\documentclass[letters]{jpsj2}

\def\runtitle{Ionic Ordering of Spinel Compound CuIr$_{2}$S$_{4}$}
\def\runauthor{Takao \textsc{Furubayashi}, Shoichi \textsc{Nagata} and Takehiko \textsc{Matsumoto}}

\title{%
Ionic Ordering in Thiospinel CuIr$_{2}$S$_{4}$
}

\author{%
Takao \textsc{Furubayashi}\thanks{furubayashi.takao@nims.go.jp}, 
Takehiko \textsc{Matsumoto} 
and Shoichi \textsc{Nagata}$^{1}$
}

\inst{%
National Institute for Materials Science, Tsukuba 305-0047\\
$^1$Department of Materials Science and Engineering, Muroran Institute of Technology, Muroran 050-8585}

\recdate{ \hspace{50mm}}

\abst{%
Crystal structure of spinel compound CuIr$_{2}$S$_{4}$ was examined by powder X-ray diffraction for the
insulating phase below the metal-insulator transition at $T_{\textrm{MI}}$ = 230 K. 
The superstructure spots are reproduced by considering the displacement of Ir atoms. 
A model for the ionic ordering of Ir$^{4+}$ and Ir$^{3+}$ with the same number is proposed for the
insulating phase on the basis of the structural analysis. 
The model suggests that the structural change at $T_{\textrm{MI}}$ is driven by the formation Ir$^{4+}$
dimers. In addition, we found that the superstructure spots becomes significantly weak below 60 K,
without any significant effects on electric and magnetic properties. 
Possible mechanism for the transition is discussed. 
}

\kword{%
spinel, metal-insulator transition, ionic ordering, dimer
}

\begin{document}
\sloppy
\maketitle

Chalcogenide spinel compound CuIr$_{2}$S$_{4}$ has the normal spinel structure with Cu in the $A$ site, 
which is tetrahedrally coordinated with S, and Ir in the $B$ site, which is octahedrally coordinated. 
The compound is metallic around room temperature and 
exhibits a transition to an insulating phase at the temperature $T_{\textrm{MI}}$ = 230 K. ~\cite{rf:1,rf:2}  
The resistivity increases by more than three decades at the transition and becomes 
semiconducting below $T_{\textrm{MI}}$.  Measurements of magnetic susceptibility 
shows that the compound is Pauli paramagnetic in the metallic phase. 
The susceptibility decreases at the transition and becomes diamagnetic, indicating the 
disappearance of the Fermi surface at the transition.
The transition is accompanied by the change of the crystal structure.
The powder x-ray diffraction pattern of the low-temperature phase was found to be 
expressed approximately by the tetragonal structure, obtained by expanding the cubic 
unit cell of spinel along one of the unit vector and shortening along the other two slightly. ~\cite{rf:2}
In addition, some extra reflections that cannot be explained by the assumed tetragonal structure 
were observed in the diffraction pattern. It has been found by recent investigations that 
the lattice is slightly distorted to be triclinic. ~\cite{rf:3} Electron diffraction studies have 
shown that the superstructure spots are indexed by using the triclinic lattice. ~\cite{rf:4}  
Neutron powder diffraction shows that 
the superstructure spots are mostly explained by the triclinic structure. ~\cite{rf:5}
It is also to be noticed that, in the electron diffraction patterns, the superstructure spots observed below $T_{\textrm{MI}}$ almost 
disappears by decreasing temperature below 50 K. ~\cite{rf:4}
Interestingly, the application of high pressures stabilizes the insulating phase ~\cite{rf:2, rf:6} on the contrary to most of other
materials that exhibit metal-insulator transition.  

The mechanism of the transition is still unclear.
Many works have been done for examining the electronic state of the material since 
the discovery of the metal-insulator transition. Results of Cu NMR, ~\cite{rf:7} XPS ~\cite{rf:8} and 
band calculations ~\cite{rf:9} suggest that of Cu atom has almost filled 3d orbit resulting in 
the Cu$^{1+}$ state. Ir M\"{o}ssbauer effect was measured for determining the ionic state of Ir.
~\cite{rf:10} 
The result seems to indicate the presence of two kinds of ionic states. However, the conclusion is not yet
convincing. 
Anyway, it seems most probable that the ionic configuration of 
Cu$^{1+}$Ir$^{3+}$Ir$^{4+}$S$^{2-}_{4}$ and the ordering of two anions, Ir$^{3+}$ 
and Ir$^{4+}$ is realized as the insulating state. In addition, magnetization measurements and NMR indicates that
the insulating phase is non magnetic. The Ir$^{4+}$ ion is expected to have a spin of $S$=1/2 with the electronic configuration of $(5d\epsilon )^5$, while Ir$^{3+}$ is in the state $S$=0 with $(5d\epsilon )^6$. Therefore, the  Ir$^{4+}$ possibly forms dimers resulting 
in the spin singlet state. 

This paper describes extended analysis of the powder x-ray diffraction of the low-temperature 
phase of CuIr$_{2}$S$_{4}$. Positions of Ir atoms in the unit cell were obtained from the diffraction profiles
including the superstructure reflections.
A model for the ionic ordering of Ir$^{3+}$ and Ir$^{4+}$ is proposed for the insulating phase.
The deformation from cubic to nearly tetragonal at $T_{MI}$ is consistently understood by the ionic ordering
model. We show that the refined structure is consistent with the assumption that Ir$^{4+}$ ions form dimers. 
In addition, we found that further structural change occurs below 60 K, as indicated by the electron diffraction. ~\cite{rf:4}
Possible mechanism for this is discussed.

Samples were prepared in the solid reaction method described previously. ~\cite{rf:1,rf:2}
Powder X-ray diffraction was measured by using a conventional diffractometer equipped 
with a curved-graphite monochromator on the counter side. A Cu-K$\alpha$ radiation was 
used for the measurements. Low temperatures down to 9 K were attained by a 
closed-cycle helium refrigerator. The sample powder was pressed onto a copper 
plate coated slightly with Apiezon N grease and the copper plate was attached 
to the cold part of the refrigerator. Lattice parameters were determined from the refinement 
of the diffraction patterns by using the RIETAN Rietveld analysis program. ~\cite{rf:11,rf:12}

The diffraction pattern at room temperature was well reproduced by the normal spinel structure with 
the space goup $Fd\bar{3}m$ and the lattice constants, $a$=9.8536 \AA \  and $u$=0.387.
Figure 1 shows the X-ray diffraction patterns at temperatures below $T_{MI}$. 
As reported previously, ~\cite{rf:2} the structure below $T_{MI}$ is approximately expressed by the
tetragonal structure obtained by expanding the cubic 
lattice along [001] direction and shortening along [100] and [010]. 
In addition, we observe some extra spots that can not be explained just by deforming the unit cell. 
The extra spots are indexed by integers or half-integers as shown for the pattern at 120 K in Fig.1. 
This means that the unit length along the principal axes should be at least double of each original length. 
Such spots were also observed in the electron diffraction patterns. ~\cite{rf:4}
We also found that such extra spots become weaker at still lower temperatures. The pattern at 9K, recorded just after cooled 
from room temperature with the cooling time of 1.5 hours, contains the extra spots, although slightly weaker than 
at 120K. The extra spots become significantly weak after kept at 9 K for 12 hours. 
This observation corresponds to the results of electron diffraction studies reported by Sun \textit{et al.}. ~\cite{rf:4}
They showed that the extra spots observed below $T_{MI}$ become weak when cooled below 50 K. 
With increasing the temperature, the extra spots recovers at about 60 K as sown in Fig. 1. 
Thus, the transformation of the crystal structure occurs at 60 K with a large thermal hysteresis. 
In the observation by electron diffraction, the extra spots become weak  as soon as cooled below 50 K. ~\cite{rf:4} 
Thus, the hysteresis is not so large.
This is presumably because the irradiation by electron beam is favorable for the transformation to the
lower-temperature phase. 
We tried to find out some anomalies in electronic and magnetic properties accompanied by the structural
transformation at 60 K. 
Electric resistivity and magnetic susceptibility were measured after waiting for more than 12 hours at 9K. 
The results were compared with those during the cooling process and indicated no appreciable difference. 
Thus, the transformation at 60 K has no apparent effect on the electronic state of CuIr$_{2}$S$_{4}$.

First, we discuss the structure showing the pattern with extra spots. The pattern at 120 K in Fig.1 was
analyzed by using a triclinic unit cell ~\cite{rf:3, rf:4}. The unit
vectors are expressed by,

\[\vec{a}_{\textrm{t}}=-\vec{b}_{\textrm{d}}+\vec{c}_{\textrm{d}}\]
\begin{equation}
\vec{b}_{\textrm{t}}=(2\vec{a}_{\textrm{d}}+\vec{b}_{\textrm{d}}-\vec{c}_{\textrm{d}})/2
\end{equation}
\[\vec{c}_{\textrm{t}}=(\vec{b}_{\textrm{d}}+\vec{c}_{\textrm{d}})/2,\]

\noindent where the suffix t indicates the triclinic cell and the suffix d corresponds to the unit cell obtained by
deforming the cubic cell of spinel. Figure 2 shows the relation of each unit cell.
It can be shown that all the reflections, including the superstructure spots, can be indexed with integers by this
tetragonal unit cell. ~\cite{rf:3, rf:4}   
We did the analysis in the following process.
First the pattern was fitted by using the tetragonal structure, obtained by deforming the cubic unit cell as
assumed in the previous work.
Then the lattice constants and the atomic position of Ir atoms was optimized by using the triclinic unit cell
defined by Eq. 1.
The space group $P\bar{1}$ with the inversion symmetry was assumed.
The positional parameters for Cu and S and the thermal factors were fixed to the values obtained from the
tetragonal model 
because the parameters became too many to be determined reasonably.
The obtained results are shown in Table I. It was found that the diffraction pattern is reproduced by the
calculation as shown in Fig. 3.
Then we assigned the ionic species of Ir from the viewpoint of the distance between neighboring Ir atoms, shown in Table II.
The distance is 2.90 \AA \ between the sites Ir1 and Ir4 and 3.01 \AA \ between Ir2 and Ir3. These are
significantly shorter than the value 3.43 \AA \ when equally spaced along the [110]$_{\textrm{d}}$ direction. These two pairs of
Ir atoms are closer than any other combinations, as shown in Table 2.
Therefore, these 4 sites are assigned to Ir$^{4+}$ by considering that the distance becomes short in forming a dimer.

The obtained ionic configuration is indicated in Fig.4.
We find two kinds of ionic configuration appear on the framework of the deformed cubic unit cell. 
The two patterns, denoted as A and B in the figure, are piled with each other along the $a_{\textrm{d}}$, $b_{\textrm{d}}$ and $c_{\textrm{d}}$ axes,
forming the $2\times2\times2$ cycle in the cubic unit cell. Thus, the configuration is consistent with the
half-integer indices in the diffraction.
The characteristic of the configuration is that pairs of Ir$^{4+}$ and Ir$^{3+}$ appear alternately in all the
Ir chains along the $[110]_{\textrm{d}}$ and $[1\bar{1}0]_{\textrm{d}}$ directions. 
The distance between two Ir$^{4+}$ ions in these directions in the (001)$_{\textrm{d}}$ plane becomes substantially smaller than any other
combinations. 
Thus, the present model is consistent with the deformation to the nearly tetragonal structure obtained by shortening the lattice constants in the (001)$_{\textrm{d}}$ plane. 
It is reasonable to consider that the structural deformation is driven by the formation of the Ir$^{4+}$ dimers.
It is to be noted, however, that the present model violates the Anderson's theorem ~\cite{rf:13} considered for the ionic ordering in Fe$_{3}$O$_{4}$ with the same spinel structure.
According to the theorem, every tetrahedra formed by Ir should be separated to two Ir$^{4+}$ and two
Ir$^{3+}$ 
for minimizing the energy of the Coulomb interaction. 
As shown in Fig.4, however, the present model contains some tetrahedra with a 3 to 1 combination of two ions. 
This is presumably because the present model is favorable for the energy of lattice distortion 
despite that it is unfavorable in view of Coulomb interaction. 
However, more detailed studies would be required for clarifying the problem.

Another experimental result to be discussed is the change of the structure at 60 K. 
Superstructure spots become significantly weak, although the main reflections, determining the form of the
unit cell, do not change, as shown in Fig. 1.
We have shown that such superstructure spots appears to originate from the formation of Ir$^{4+}$ dimers. 
Thus, the structural change at 60 K would be related to the configuration of the ionic ordering. 
On the other hand, we observed no significant change in electric and magnetic properties around 60 K. 
The sample is still insulating under the temperature of the structural change, the disappearance of the
superstructure spots.
Thus, it does not seem to occur that the dimer state of Ir$^{4+}$, formed at $T_{\textrm{MI}}$=230 K, is destroyed
again at this temperature. 
One possible explanation is that the charge ordering becomes incommensurate to the lattice. 
Above 60 K, two ions are ordered as shown in Fig.4 with the commensurate relation with the lattice. 
The position of the dimers may be disordered at low temperatures, resulting in the diffuse and weak
superstructure reflections. 
It seems strange, however, that the spatial disorder increases with decreasing the temperature. 
For clarifying this problem, more detailed investigations would be required both theoretically and experimentally.

In summary, a model for the ionic ordering of Ir$^{3+}$ and Ir$^{4+}$ for the insulating phase of
CuIr$_{2}$S$_{4}$ below $T_{MI}$ was proposed on the basis of the analysis of the superstructure spots in
the powder X-ray diffraction pattern. 
In this model, each pair of Ir$^{4+}$ and Ir$^{3+}$ appears alternatively along [110]$_{\textrm{d}}$ and [1$\bar{1}$0]$_{\textrm{d}}$ directions in the deformed cubic unit cell. 
The deformation from cubic to nearly tetragonal structure is reasonably
understood as driven by the formation of Ir$^{4+}$ dimers in the (001)$_{\textrm{d}}$ plane. 
With decreasing the temperature further, the superstructure spots almost disappears below 60 K. 
This suggests another structural transition, 
although no appreciable change was observed in electric and magnetic properties. 

Very recently, Ragaelli \textit{et al.} proposed a ordering model of Ir$^{3+}$ and Ir$^{4+}$ for CuIr$_{2}$S$_{4}$ based on the analysis of X-ray and neutron diffraction. Their result of the ordering patten of Ir$^{3+}$ and Ir$^{4+}$ is the same as presented in the present work, although a different triclinic unit cell is adopted.

\begin{table}

\caption{chrystalographic parameters of CuIr$_{2}$S$_{4}$ at 120 K}
\label{t1}
Space Group: $P\bar{1}$ \\
$a_{\textrm{t}}$=13.975, $b_{\textrm{t}}$=11.959, $c_{\textrm{t}}$= 6.988 (\AA), $\alpha_{\textrm{t}}$= 91.11, $\beta_{\textrm{t}}$= 88.19, $\gamma_{\textrm{t}}$= 125.77
(deg)\\ 
($a_{\textrm{d}}$=9.704, $b_{\textrm{d}}$=9.725, $c_{\textrm{d}}$= 10.037 (\AA), $\alpha_{\textrm{d}}$= 90.00, $\beta_{\textrm{d}}$= 90.06, $\gamma_{\textrm{d}}$= 90.04
(deg))\\
$B$(Cu)= 0.309, $B$(Ir)= 0.321, $B$(S)= 0.468 (\AA $^{2}$)\\
$R_{\textrm{wp}}$ = 13.01 \%,    $R_{\textrm{p}}$ = 10.11 \%, $R_{\textrm{R}}$ = 15.52 \% \\
All atoms are in the 2$i$ site.
{\small \begin{halftabular}{@{\hspace{\tabcolsep}\extracolsep{\fill}}cccc} \hline
site & $x$ & $y$ & $z$  \\ \hline
Cu1 &  0.5625 & 0.625 & 0.25 \\
Cu2 &  0.9375 & 0.875 & 0.25   \\
Cu3 &  0.0625 & 0.625&  0.25    \\
Cu4 &  0.4375 & 0.875 & 0.25   \\
Ir1(4+) & 0.251(4) & 0.019(7) & 0.019(5) \\
Ir2(4+) & 0.238(5) & 0.482(7) & 0.005(8) \\
Ir3(4+) & 0.015(5) & 0.261(5) & 0.216(6)   \\
Ir4(4+) & 0.247(4) & 0.227(4) &  0.231(8)   \\
Ir5(3+) &  0.516(3)  & 0.263(5) & 0.245(6)   \\
Ir6(3+) & 0.757(4) & 0.254(4) & 0.239(7)   \\
Ir7(3+) &  0.751(3) & -0.009(4) & 0.484(5)   \\
Ir8(3+) &   0.755(5) & 0.502(5) & 0.496(8)    \\
S1&  0.8787 & 0.2457 & 0.0202   \\
S2&  0.3798 & 0.2457&  0.0224   \\
S3&  0.1341 & 0.2543&  0.0224   \\
S4&  0.6341 & 0.2543 & 0.0224   \\
S5&  0.3659 & 0.0043&  0.2361   \\
S6&  0.8659 & 0.0043 & 0.2361  \\ 
S7& 0.3617 & 0.4957 & 0.2361 \\
S8& 0.8617 & 0.4957&  0.2361   \\
S9&  0.1373 & 0.0043 & 0.2617  \\ 
S10&  0.6383 & 0.0043 & 0.2639   \\
S11&  0.1341 & 0.4957  &0.2639  \\
S12&  0.6341 & 0.4957&  0.2639  \\
S13&  0.3659 & 0.2457 & 0.4776  \\
S14&  0.8659 & 0.2457 & 0.4776  \\
S15&  0.1202 & 0.2543 & 0.4776 \\
S16&  0.6202&  0.2543 & 0.4776 \\
\hline
\end{halftabular}
}\end{table}

\newpage

\begin{table}
\caption{distance of neighboring Ir sites at 120 K}
\label{t2}
\begin{halftabular}{@{\hspace{\tabcolsep}\extracolsep{\fill}}cccc} \hline
site & distance(\AA) &  &   \\ \hline
Ir$^{4+}$-Ir$^{4+}$ &   & Ir$^{3+}$-Ir$^{4+}$ & \\
1-3 &  3.66 & 1-5 & 3.50  \\
1-4 &  2.90 & 1-6 & 3.65  \\
2-3 & 3.01  & 1-7 & 3.48, 3.52\\
2-4 & 3.53 & 2-5 & 3.42 \\
3-4 & 3.49 & 2-6 & 3.58 \\
Ir$^{3+}$-Ir$^{3+}$ & & 2-8 & 3.49, 3.51 \\
5-6 & 3.42 & 3-6 & 3.57 \\
5-7 & 3.64 & 3-7 & 3.66 \\
5-8 & 3.38 & 3-8 & 3.45 \\
6-7 & 3.57 & 4-5 & 3.53 \\
6-8 & 3.44 & 4-7 & 3.53 \\
 & & 4-8 & 3.74 \\
\hline
\end{halftabular}
\end{table}

\newpage

\begin{figure}[tbp]
  \begin{center}
    \includegraphics[keepaspectratio=true,height=120mm]{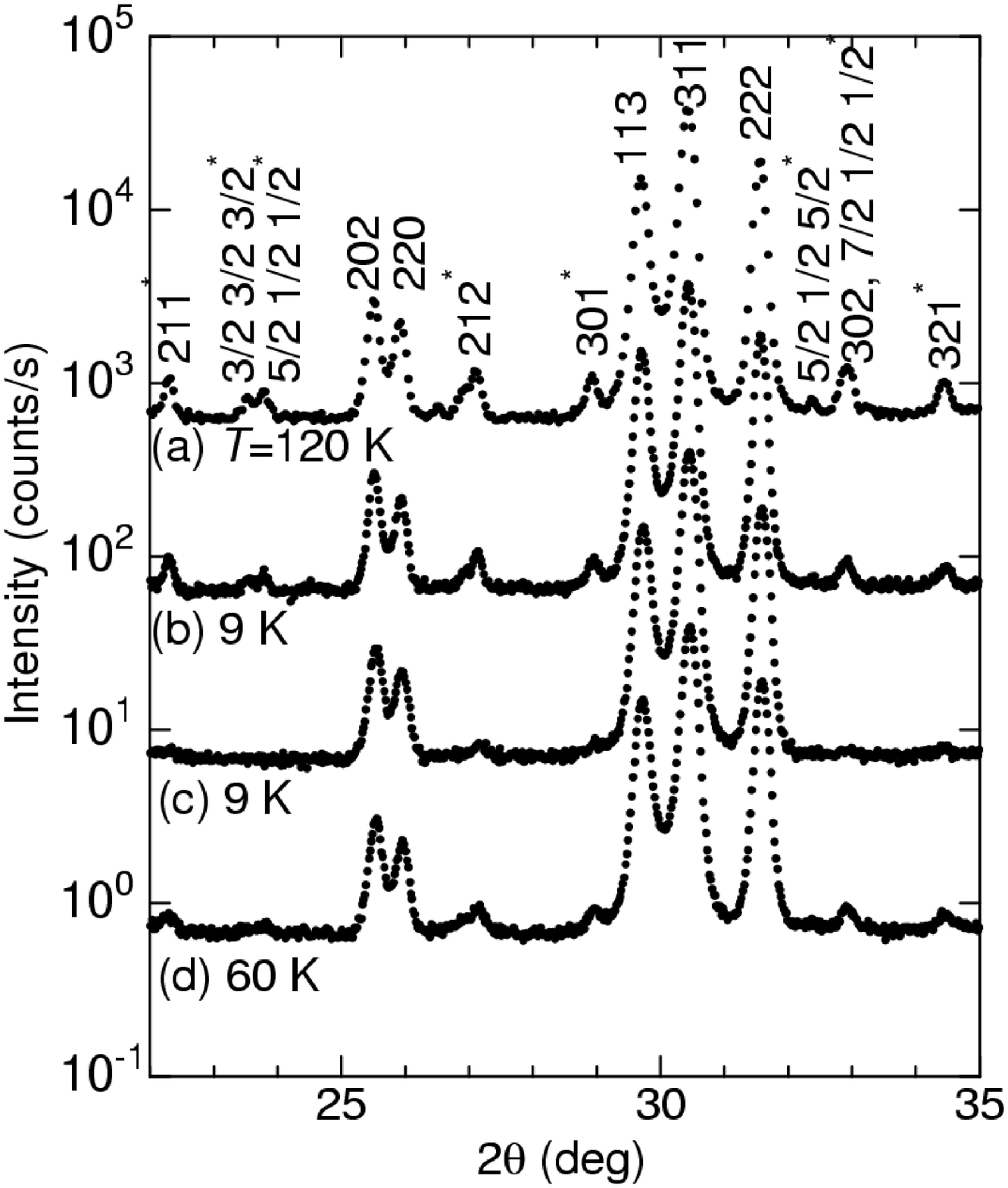}
  \end{center}
  \caption{Powder X-ray diffraction at each temperature: (a) after cooled to 120 K from room temperature, 
(b) just after cooled to 9K, (c) after waiting at 9 K for 12 hours, and (d) taken by warming to 60 K after (c). 
The indices are for the unit cell obtained by
deforming the cubic cell above $T_{\textrm{MI}}$. The indices marked by asterisks are extra
reflection, which cannot be explained by just deforming the unit cell.}
  \label{fig:figure1.eps}
\end{figure}

\begin{figure}[tbp]
  \begin{center}
    \includegraphics[keepaspectratio=true,height=60mm]{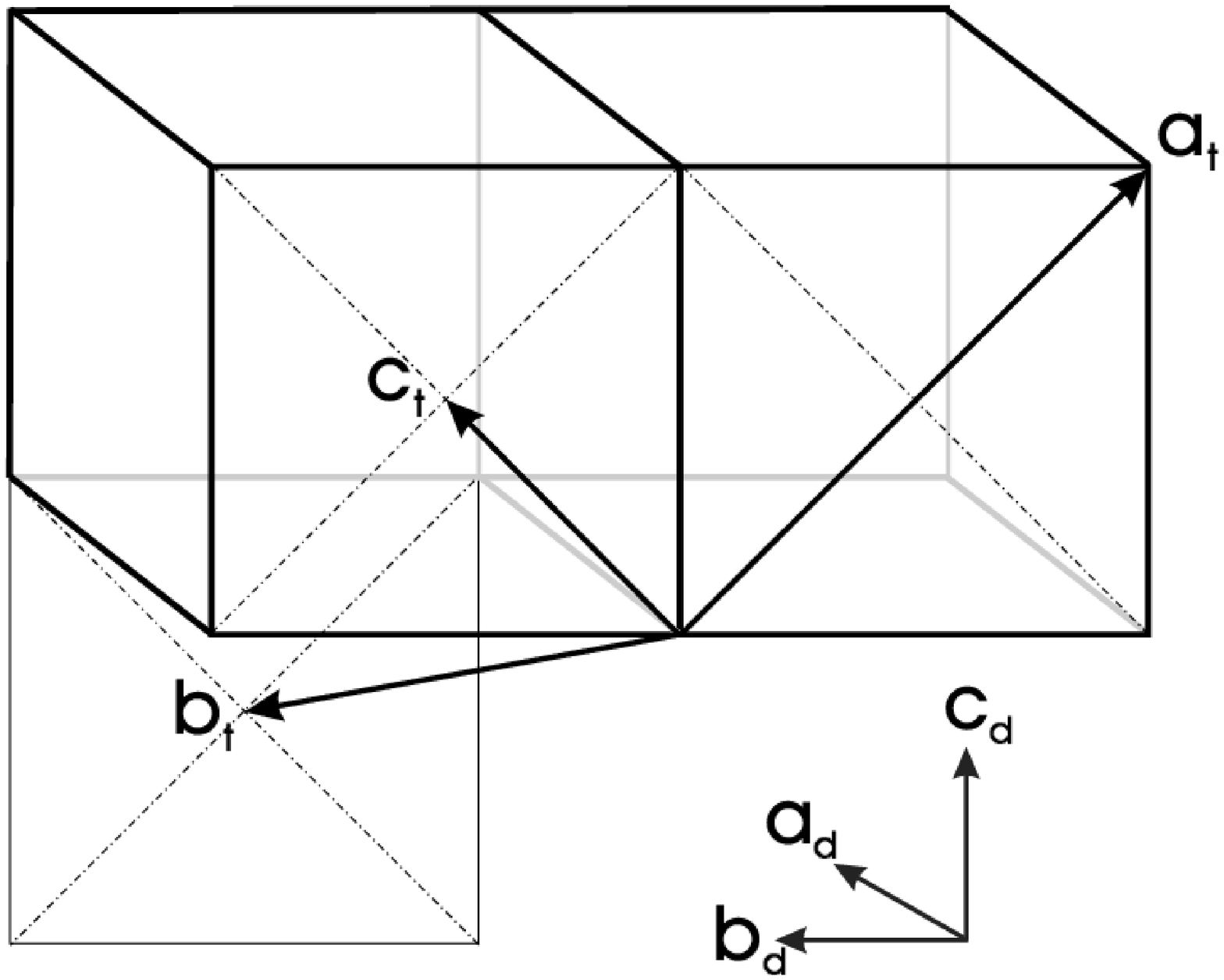}
  \end{center}
  \caption{The relation between deformed cubic (d) and triclinic (t) unit cell used in the analysis.}
  \label{fig:figure2.eps}
\end{figure}

\begin{figure}[tbp]
  \begin{center}
    \includegraphics[keepaspectratio=true,height=180mm]{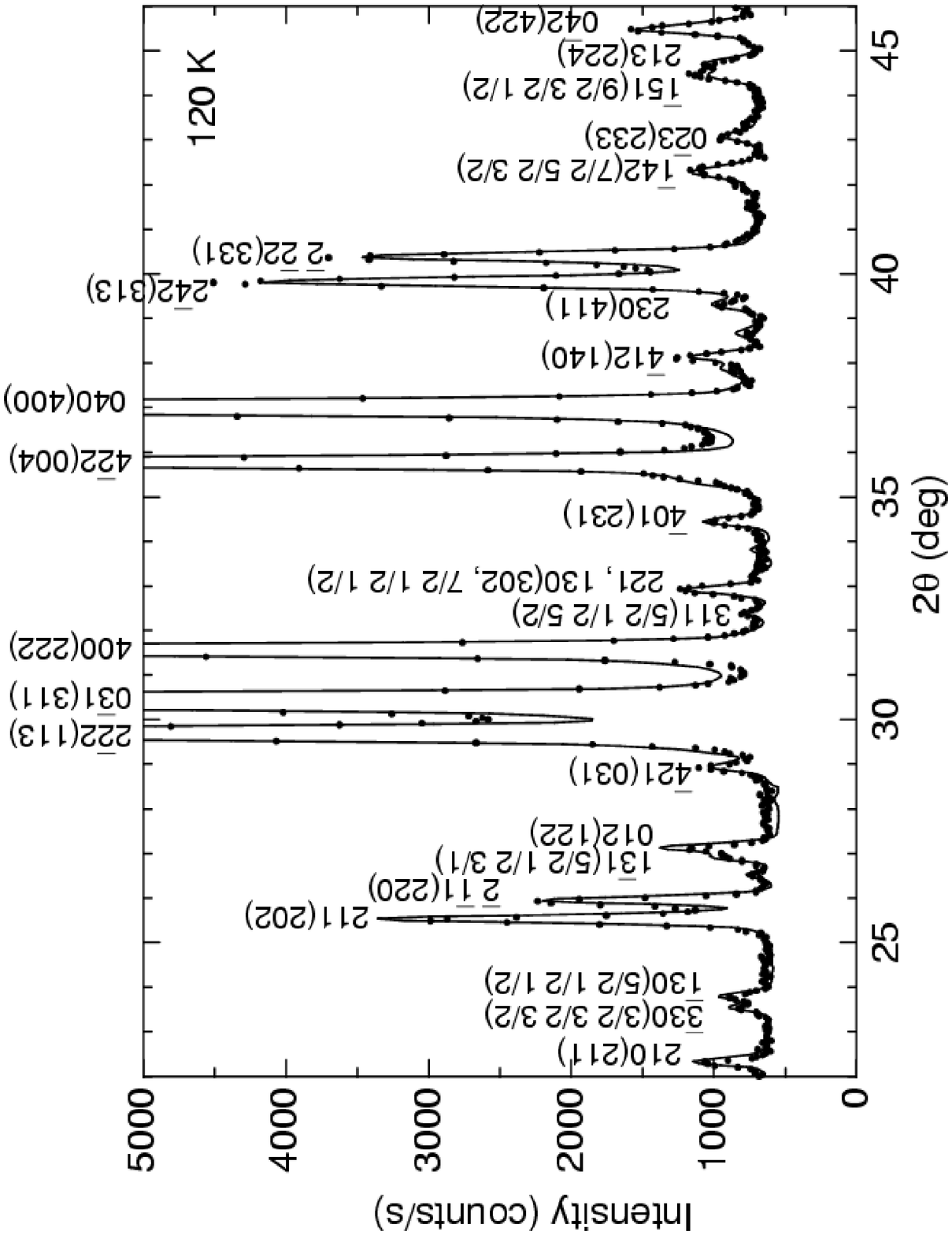}
  \end{center}
  \caption{Diffraction pattern at 120 K. The dots are the measured values and the line is the calculated curve. 
The indices (not all) are shown for the triclinic unit cell. The indices for the deformed cubic cell are in the parentheses.
}
  \label{fig:figure3.eps}
\end{figure}

\begin{figure}[tbp]
  \begin{center}
    \includegraphics[keepaspectratio=true,height=80mm]{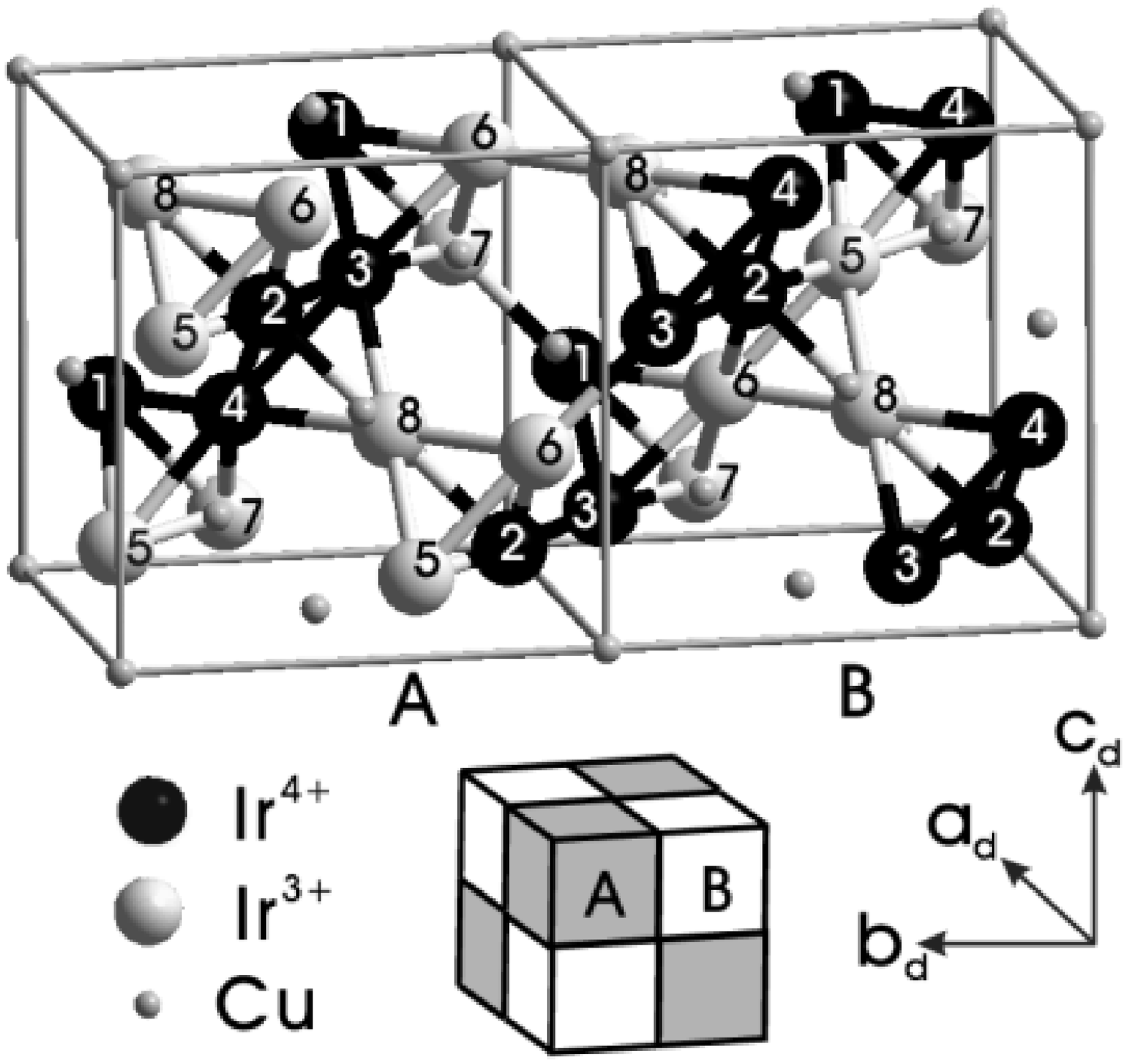}
  \end{center}
  \caption{Model of the ionic configuration of the insulating phase of CuIr$_{2}$S$_{4}$ shown for the deformed cubic cell.
The S atoms are omitted.
Two configurations, A and B, appears alternatively as shown underneath. Numbers on Ir are the site numbers shown in Table 1.}
  \label{fig:figure4.eps}
\end{figure}

\end{document}